\documentclass[12pt]{iopart}
\usepackage{iopams}
\usepackage{graphicx}

\begin{document}
\title{Efficiency of the SQUID Ratchet Driven by External Current}
%
\author{J. Spiechowicz$^1$ and J. \L uczka$^{1,2}$}
\address{$^1$ Institute of Physics, University of Silesia, 40-007 Katowice, Poland}
\address{$^2$ Silesian Center for Education and Interdisciplinary Research, University of Silesia, 41-500 Chorz{\'o}w, Poland}
\begin{abstract}
We study theoretically the efficiency of an asymmetric superconducting quantum interference device (SQUID) which is constructed as a loop with three capacitively and resistively shunted Josephson junctions. Two junctions are placed in series in one arm and the remaining one is located in the other arm. The SQUID is threaded by an external magnetic flux and driven by an external current of both constant (dc) and time periodic (ac) components. This system acts as a nonequilibrium ratchet for the dc voltage across the SQUID with the external current as a source of energy. We analyze the power delivered by the external current and find that it strongly depends on thermal noise and the external magnetic flux. We explore a space of the system parameters to reveal a set for which the SQUID efficiency is {\it globally maximal}. We detect the intriguing feature of the thermal noise enhanced efficiency and show how the efficiency of the device can be tuned by tailoring the external magnetic flux.
\end{abstract}
\submitto{\NJP}
\maketitle
\section{Introduction}
The SQUID is the most sensitive instrument which is capable of detecting and measuring even extremely small magnetic fields. It has been used successfully not only for magnetometry but also for voltage and current measurements. Its applications go far beyond the research laboratories often into commercial apparatus exploited in metrology, geophysics and medicine, see the reviews \cite{fagaly2006,matti1993}. The SQUID has been the topic of various extensive theoretical and experimental studies. Yet, a number of open problems of this setup still remain to be resolved. A prominent example may be the efficiency of the SQUID as a thermodynamical machine converting the input energy into its other forms. It is the subject of this paper. 

We study an asymmetric SQUID driven by an external current and analyze the charge transport and voltage induced across the device. The asymmetric SQUID is modeled as a ratchet far from equilibrium, i.e. as a classical Brownian particle moving in a spatially periodic potential with broken reflection symmetry and driven by a time-dependent force. In this mechanical analogy, the voltage across the SQUID corresponds to the particle velocity. The most basic measure for characterizing the motion of the Brownian particle is its long time average velocity $\langle v \rangle$. However, alone it does not give any information on quality of transport process. Is it effective or ineffective? To answer this question, we need to consider its other attributes. One of them are \emph{fluctuations} of the velocity around its average value which in the long time regime are represented by the variance $\sigma_v^2 = \langle v^2 \rangle - \langle v \rangle^2$. Then, typically the instantaneous velocity $v(t)$ takes values within the interval of standard deviation, $v(t) \in \left[\langle v \rangle - \sigma_v, \langle v \rangle + \sigma_v\right]$. Note that if fluctuations are large, i.e. if $\sigma_v > |\langle v \rangle|$, then it is possible that the particle moves for some time in the direction opposite to its average velocity $\langle v \rangle$, spread of velocities is large and overall transport is not effective. The next feature which is important in answering the question about the quality of transport phenomenon is related to the ratio of energy input into the system and its energetic output. How much of the energy input is converted into directed motion of the particle and how much of it is wasted by spreading out into environment and dissipated as heat? A proper quantifier to characterize this aspect of transport is the \emph{efficiency} of the system.

By using the correspondence between the SQUID and the mechanical ratchet system, we study three measures for evaluation of transport quality: average voltage, its fluctuations and the efficiency of the SQUID. In the previous paper \cite{spiechowicz2014} we analyzed the average voltage in this setup for wide parameter regimes: covering the overdamped to moderate damping regime up to its fully underdamped regime. We found the intriguing features of a negative absolute and differential conductance, repeated voltage reversals, noise induced voltage reversals and solely thermal noise-induced ratchet voltage. We identified a set of parameters for which the ratchet effect is most pronounced and showed how the direction of transport can be controlled by tailoring the external magnetic flux. The main emphasis of that work laid on formulating and exploring conditions that are necessary for the generation and control of transport \cite{hanggi2005,hanggi2009}, its direction, magnitude as well as its dependence on system parameters. However, apart from these well investigated questions other important features concerning the quality of transport \cite{machura2004, machura2005, machura2006} have remained unanswered. Therefore in this paper we concentrate on this topic and connection between the directed transport expressed in terms of the dc voltage, its fluctuation characteristics and energetics of the SQUID.

Theoretical aspects  considered in the  paper concern not only our specific SQUID ratchet but a much wider class of systems and problems. There are many experiments on a number of ratchet systems \cite{Reimann}, in particular superconducting ratchets \cite{sterck2005}, a part of which can be controlled by an external magnetic field \cite{villegas2003,togawa2005,cole2006,ooi2007} as well as  theoretical studies of such systems driven by harmonic and biharmonic external currents \cite{savelev2004,savelev2004epj,savelev2004epl,zhu2004,zhu2003}. However, the efficiency of transport has not been analyzed in the above-cited papers.

The structure of the paper is as follows. In Sec. II, we recall the model of a SQUID rocking ratchet which is composed of three resistively and capacitively shunted Josephson junctions. In Sec. III we define mean values of arbitrary state functions in the long time regime. Then in Secs. IV-VI, the quantities characterizing the quality of the transport such as the voltage fluctuations, the energy balance and the (Stokes) efficiency are introduced, respectively. In Sec. \ref{results} we elaborate on key aspects of transport efficiency in the system: starting from the power delivered by the externally applied current, covering the tailoring of the Stokes efficiency of the device, up to presentation of the regime for which thermal noise enhances the efficiency and discussion about the impact of variation of the external magnetic flux on the efficiency of the SQUID. Finally, the last section provides a summary.
\begin{figure}[t]
    \centering
    \includegraphics[width=0.49\linewidth]{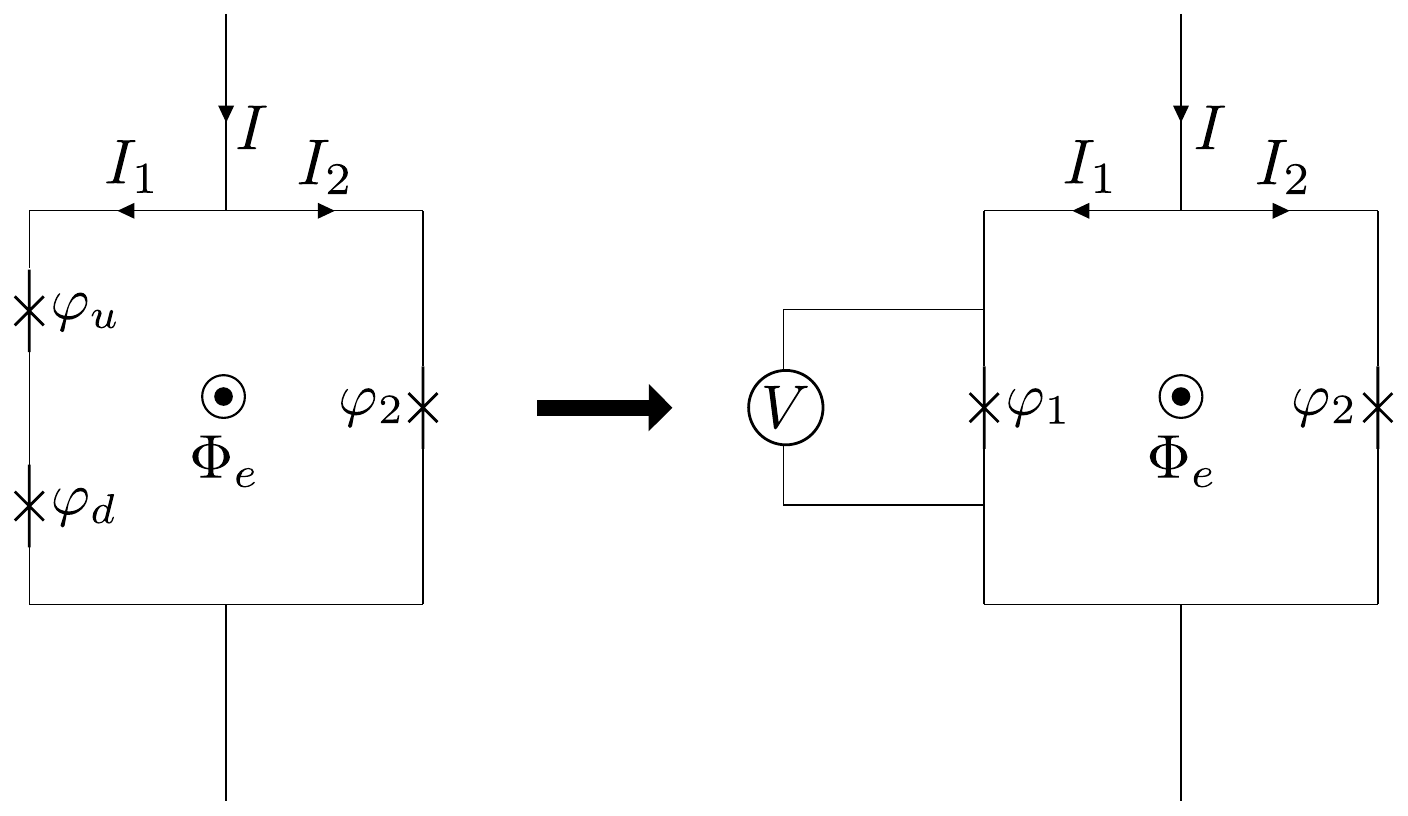}
 \caption{Schematic asymmetric SQUID composed of three Josephson junctions and the equivalent circuit composed of two junctions. The Josephson phase difference is $\varphi_1 = \varphi_u + \varphi_d$, the externally applied current is $I$, the current through the left and right arms is $I_1$ and $I_2$, respectively. The external magnetic flux is $\Phi_e$ and the instantaneous voltage across the SQUID is $V=V(t)$. The long time average voltage $\langle V \rangle$ across the SQUID is expressed by the relation $\langle V \rangle = \hbar \langle \dot{\varphi}_1 \rangle/2e = \hbar \langle \dot{\varphi}_2 \rangle/2e$.}
    \label{fig1}
\end{figure}
\section{Model of the SQUID ratchet}
The asymmetric SQUID \cite{sterck2005, weiss2000, sterck2002, berger2004,  sterck2009} is presented in Fig. 1. It is a loop with two resistively and capacitively shunted Josephson junctions \cite{josephson} in the left arm and one in the right arm. The crosses denote the junctions and $\varphi_k \equiv \varphi_k(t)$ ($k=u, d, 1, 2$) are the phase differences across them. Each junction is characterized by the capacitance $C_k$, resistance $R_k$ and critical Josephson supercurrent $J_k$, respectively. 

The SQUID is threaded by an external magnetic flux $\Phi_e$ and driven by an external current $I = I(t)$ which is composed of the static dc current $I_0$ and the ac component of amplitude $A$ and angular frequency $\Omega$, namely
\begin{equation}
    \label{eq14}
    I(t) = I_0 + A \cos(\Omega t).
\end{equation}
To reduce a number of parameters of the model, we consider a special case when two junctions in the left arm are identical, i.e. $J_u = J_d \equiv J_1, R_u = R_d \equiv R_1/2, C_u = C_d \equiv 2C_1$. In some regimes \cite{spiechowicz2014}, two junctions in a series can be considered as one for which the supercurrent-phase relation takes the form $J_1 \sin{\left(\varphi_1/2\right)}$, where $\varphi_1 = \varphi_u + \varphi_d$. This result is also derived in Ref. \cite{zapata1996prb} for an effective double-well structure described in terms of a double-barrier potential (cf. Eq. (23) therein).

The total magnetic flux $\Phi$ piercing the loop is a sum of the external flux $\Phi_e$ and the flux due to the flow of currents,
\begin{equation}
    \label{eq8}
    \Phi = \Phi_e + L i, 
\end{equation}
where $L$ is the loop inductance and $i\equiv i(t)$ is the circulating current which tends to screen the magnetic flux. In the "dispersive" operating mode of the SQUID \cite{baronepaterno}, i.e. when the condition $|Li| \ll \Phi_{0}$ holds true ($\Phi_0 = h/2e$ is the flux quantum), the phase $\varphi \equiv \varphi_1$ obeys the Stewart-McCumber type equation of the form \cite{spiechowicz2014} 
\begin{equation}
    \label{Lang1}
	\frac{\hbar}{2e} C \ddot{\varphi} + \frac{\hbar}{2e} \frac{1}{R} \dot{\varphi} + J(\varphi) = I(t) +	\sqrt{\frac{2k_B T}{R}}\,\xi(t),
\end{equation}
where the effective supercurrent $J(\varphi)$ reads
\begin{equation} \label{J}
    J(\varphi) = J_1\sin{\left( \frac{\varphi}{2} \right)} + J_2\sin{(\varphi + \tilde{\Phi}_e)}. 
\end{equation}
The parameters are: $C = C_1 + C_2$, $R^{-1} = R_1^{-1} + R_2^{-1}$, 
$k_B$ is the Boltzmann constant, $T$ is temperature of the system and ${\tilde{\Phi}_e}= 2\pi \Phi_e/\Phi_0$ is the dimensionless external magnetic flux. Thermal fluctuations are modeled by $\delta$-correlated Gaussian white noise $\xi(t)$ of zero mean and unit intensity
\begin{equation}
    \label{eq13}
    \langle \xi(t) \rangle = 0, \quad \langle \xi(t)\xi(s) \rangle = \delta(t-s).
\end{equation}
The Stewart-McCumber equation (\ref{Lang1}) has the form of a Langevin equation and describes a non-Markovian stochastic process for the phase $\varphi$. In the extended space $\{\varphi, \dot \varphi\}$, it has the Markovian property and all well known methods can be applied to analyze it.

Eq. (\ref{Lang1}) can be interpreted in the framework of a model of a classical Brownian particle. It helps develop the intuition and interpretation. In the one-to-one correspondence, the particle position $x$ translates to the phase $\varphi$, the particle velocity $v= \dot{x}$ to the voltage $V \propto \dot \varphi$, the conservative force to the supercurrent $J(\varphi)$, the external force to the current $I(t)$, the mass $m$ to the capacitance $m \propto C$ and the friction coefficient $\gamma$ to the normal conductance $\gamma \propto G=1/R$. It is important to note that the friction $\gamma$ is not proportional to the normal resistance $R$ (as one could expect in the case of electrical circuits) but to the inverse of $R$. The reason is that plasma oscillations of the junction are more damped if more normal electrons couple to the oscillating condensate (i.e. when $G$ is greater). The voltage accelerates normal electrons and their kinetic energy is dissipated into heat. Thus the plasma oscillations converts into heat with a rate proportional to the conductance $G$ \cite{nanowire}.
\section{Asymptotic mean values}
The main characteristics of the system are the current-voltage curves in the long time regime. It can be shown that for the external current (\ref{eq14}), they can be extracted from the relation for the averaged voltage $\langle V \rangle$ developed across the SQUID \cite{spiechowicz2014} 
\begin{equation} 
    \label{A1}
	\langle V \rangle =\frac{\hbar}{2e} \langle \dot{\varphi}_1 \rangle = \frac{\hbar}{2e} \langle \dot{\varphi}_2 \rangle, 
\end{equation}
where $\dot \varphi_1$ can be obtained from the Stewart-McCumber equation (\ref{Lang1}) for the phase $\varphi=\varphi_1$. The above relation holds true when the average is only over the period $T=2\pi/\Omega$ of the ac current. It is also valid in the long time regime when the averaging is performed over initial conditions and all realizations of thermal fluctuations. Below we present a precise definition of this procedure.

It is convenient to rewrite Eq. (\ref{Lang1}) for the phase $\varphi$ and the voltage $V \equiv V(t) = (\hbar/2e) \dot{\varphi}$ in the Ito form
\begin{eqnarray} 
d\varphi &=& \frac{2e}{\hbar} V dt, \label{Ito1}\\
dV &=& \frac{1}{C}\left[-\frac{1}{R} V - J(\varphi) +I(t)\right] dt
+ \frac{1}{C} \sqrt{\frac{2k_B T}{R}}\,dW(t), \label{Ito2}
\end{eqnarray} 
where $dW(t) =\xi(t) dt$ is the differential of the Wiener process of zero mean and the second moment $\langle dW(t)dW(t)\rangle = dt$. The pair $\{\varphi, V\}$ form a Markovian process and its probability density $P = P(\varphi, V, t)$ obeys the Fokker-Planck equation 
\begin{eqnarray} \label{FP}
\frac{\partial P}{\partial t} = -\frac{2e}{\hbar} V \frac{\partial}{\partial \varphi} P + \frac{1}{C} \frac{\partial}{\partial V} \left [ \frac{1}{R} V + J(\varphi) - I(t)\right] P + \frac{k_B T}{R C^2} \frac{\partial^2}{\partial V^2} P
\end{eqnarray}
with the initial condition $ P(\varphi, V, 0) = p(\varphi, V)$, where a given probability density $p(\varphi, V)$ describes the initial distribution of the phase $\varphi(0)$ and voltage $V(0)$.
 
For any state function $f(\varphi, V)$, its mean value $\langle f(\varphi, V) \rangle_t$ at time $t$ is calculated from the relation
\begin{eqnarray} \label{meant}
\langle f(\varphi, V) \rangle_t = \int_0^{2\pi}d\varphi \int_{-\infty}^{\infty}dV \, f(\varphi, V) P(\varphi, V, t).
\end{eqnarray}
Because the system is driven by the time-periodic current $I(t)$, the probability density $P(\varphi, V, t)$ approaches for long time the asymptotic {\it periodic} form $P_{as}(\varphi, V, t)$, namely \cite{jung1990, jung1993}
\begin{eqnarray} \label{Pas}
P_{as}(\varphi, V, t) = \sum_{n=-\infty}^{\infty} W_n(\varphi, V) 
e^{in\Omega t}, 
\end{eqnarray}
where the Fourier coefficients $W_n(\varphi, V)$ are solutions of the differential equations obtained from the Fokker-Planck Eq. (\ref{FP}). 
The time-dependent asymptotic mean value 
\begin{eqnarray} \label{asym}
\langle f(\varphi, V) \rangle^{as}_t = \int_0^{2\pi}d\varphi \int_{-\infty}^{\infty}dV \, f(\varphi, V) P_{as}(\varphi, V, t)
\end{eqnarray}
is also a periodic function of time. If we are interested in its {\it time-independent} form, the time averaging over the additional period $T=2\pi/\Omega$ of the ac current has to be performed 
\begin{eqnarray} \label{mean}
\langle f(\varphi, V) \rangle &=& \frac{\Omega}{2\pi} \int_t^{t+2\pi/\Omega} \langle f(\varphi, V) \rangle_{u}^{as} \, du \nonumber\\
&=& \lim_{t\to\infty} \frac{\Omega}{2\pi} \int_t^{t+2\pi/\Omega} 
\langle f(\varphi, V) \rangle_u \, du.
\end{eqnarray}
In a particular case, when $f(\varphi, V) = V$, we get the time-independent asymptotic mean voltage $\langle V \rangle$. Similarly, when $f(\varphi, V) = V^k (k=2, 3, ...)$, we obtain the stationary statistical moments of the voltage $\langle V^k \rangle$.
\section{Fluctuations of voltage}
The asymptotic average voltage $\langle V \rangle$ calculated according to the prescription (\ref{mean}) is the most important transport characteristics of the system. The magnitude of the instantaneous voltage $V(t)$ can be much larger than its mean value. Moreover, the fluctuations of voltage in the long time regime can also be large. They are described by the voltage variance
\begin{equation}
    \label{eq6}
    \sigma^2_V = \langle V^2 \rangle - \langle V \rangle^2.
\end{equation}
The voltage typically ranges within the interval of several standard deviations
\begin{equation}
    \label{eq7}
    V(t) \in [ \langle V \rangle - n\sigma_V, \langle V \rangle + n \sigma_V ], \quad n=1, 2,...
\end{equation}
If the standard deviation $\sigma_V$ is large, i.e. when $\sigma_V > |\langle V \rangle|$, the voltage $V(t)$ can spread far from its average value and even assume the sign opposite to it. It is a case for protein motors in biological cells where the instantaneous velocity changes direction very rapidly and its absolute value is several orders of magnitude larger than the average velocity \cite{schilwa}.
\section{Energetics of the SQUID}
The SQUID is a device which converts input energy into its other forms. It is provided by the external current $I(t)$ and the energy flow is determined by the equation of motion (\ref{Lang1}). In the mechanical interpretation, the kinetic energy of the particle corresponds to the energy stored in the system of capacitance $C$, namely
\begin{eqnarray} \label{Ec}
E_C \equiv E_C(V(t)) = \frac{1}{2} C V^2(t). 
\end{eqnarray}
The particle potential energy translates to the Josephson energy accumulated in the junction when the supercurrent flows through it 
\begin{eqnarray} \label{Ej}
E_J &\equiv& E_J(\varphi(t)) = \int_0^t J(\varphi(u)) V(u) du \nonumber\\ &=& \frac{\hbar}{2e} \int_0^t J(\varphi(u)) \dot\varphi(u) du = \frac{\hbar}{2e} \int_0^{\varphi(t)} J(\phi) d\phi \nonumber\\ &=& -\frac{\hbar}{2e} \left\{2 J_1\cos{\left[ \frac{\varphi(t)}{2} \right]} + J_2\cos[\varphi(t) + \tilde{\Phi}_e]\right\} + const.
\end{eqnarray}
The sum
\begin{eqnarray} \label{E}
E \equiv E_C + E_J
\end{eqnarray}
is the total energy of the system. Its balance can be obtained from Eqs. (\ref{Ito1})-(\ref{Ito2}). For this purpose we apply the Ito differential calculus to both functions $E_C(V)$ and $E_J(\varphi)$
\begin{eqnarray} \label{dif}
d E_C &=& \frac{dE_C}{dV} dV + \frac{1}{2} \frac{d^2E_C}{dV^2} dV dV + ... \nonumber \\ &=& \left[-\frac{1}{R} V^2 -J(\varphi) V + I(t) V + \frac{k_BT}{RC} \right] dt + \sqrt{\frac{2k_B T}{R}}\,V dW(t), \\ d E_J &=& \frac{dE_J}{d\varphi} d\varphi = J(\varphi) V dt. 
\end{eqnarray}
Next, for both sides of the above equations, we calculate the mean values. Exploiting the Ito martingale property we find the average value of the term $\langle VdW(t) \rangle_t = 0$ and obtain the energy balance equation in the form \cite{machura2004,jung2011}
\begin{eqnarray} 
\frac{d}{dt} \langle E \rangle_t &=& -\frac{1}{R} \langle V^2 \rangle_t + I(t) \langle V \rangle_t + \frac{k_BT}{RC} \label{E1}\\ &=& -\frac{1}{R} \left[\langle V^2 \rangle_t - \langle V^2 \rangle_{eq}\right] + I(t) \langle V \rangle_t, \label{E2}
\end{eqnarray} 
where $\langle \cdot \rangle_t$ denotes a mean value at time $t$ according to the prescription (\ref{meant}). In the right hand side of this equation there are three components, each of them is related to the separate process responsible for the energy change. Let us point out that the first term in Eq. (\ref{E1}) is always negative whereas the third is positive. The former describes the rate of energy loss due to dissipation and the latter refers to the energy provided by thermal equilibrium fluctuations. According to the equipartition theorem, in the thermodynamical equilibrium, when $I(t)=0$, the relation $\langle CV^2/2\rangle_{eq} = k_B T/2$ holds true. It is utilized in Eq. (\ref{E1}) to get (\ref{E2}). The second term in (\ref{E1}) characterizes the change of energy caused by the external current $I(t)$.

Now, we perform the final averaging of Eq. (\ref{E1}) according to the prescription (\ref{mean}). Since in the long time regime the average values are periodic function of time, the left hand side vanishes
\begin{eqnarray} \label{left}
\int_t^{t+2\pi/\Omega} \frac{d}{dt} \langle E \rangle_u \, du &=& \frac{C}{2} [\langle V^2(t+2\pi/\Omega)\rangle - \langle V^2(t)\rangle ] \nonumber\\ &+& [\langle E_J(\varphi(t+2\pi/\Omega))\rangle - 
 \langle E_J(\varphi(t))\rangle ] \nonumber \\&=& 0.
\end{eqnarray}
As a consequence, in the stationary regime the mean power $\mathcal P_{in}$ delivered to the system by the external current $I(t)$ over the period $T$ is expressed by the relation \cite{jung2011}
\begin{eqnarray} 
\mathcal P_{in} = \langle I(t) V \rangle = \frac{1}{R} \langle V^2 \rangle - \frac{k_BT}{RC} = \frac{1}{R} \left[\langle V^2 \rangle - \langle V^2 \rangle_{eq}\right].
 \label{e2}
\end{eqnarray}
From the above equation it follows that the amount of the energy input to the SQUID from the external driving $I(t)$ depends not only on the current itself (i.e. on $I_0, A, \Omega$) but also on properties and parameters of the device: its temperature $T$, the resistance $R$ and the capacitance $C$. In contrast, the energy supplied by thermal fluctuations does not depend on the external current but only on $T, R$ and $C$. 
\section{Efficiency of the SQUID} 
A generic definition of the efficiency of a device converting energy is a ratio between the output (work, power) and the input (power) energy. Depending on the choice of input and output, different definitions of the efficiency characterize various aspects of energy conversion in the device. To explain the problem, we use the mechanical interpretation of (\ref{Lang1}). Then the average voltage $\langle V \rangle$ corresponds to the velocity of the Brownian particle or the \emph{Brownian motor}. The thermodynamic efficiency is defined as a ratio of the work done by the motor to the energy input. If the particle is working against a constant force (load) $i_0$ then in stationary state the efficiency is defined as follows
\begin{equation}
    \label{eff1}
    \eta_{0} = \frac{ i_0 \langle V \rangle}{\mathcal P_{in}}. 
\end{equation}
In the considered case, there is no a load and the Brownian motor does not transport external objects. It works against the friction "force" 
\begin{equation}
    \label{fric}
 F_R = \frac{\hbar}{2e} \frac{1}{R} \dot{\varphi}. 
\end{equation}
(By the way, it has not the unit of Newton but if the above formula is again multiplied by the factor $\hbar/2e$ then it has the correct physical unit.) When the external force $I(t)$ is switched off, the velocity of the motor is damped to zero and the system tends to thermodynamical equilibrium. Because the motor works against the friction force, we can utilize its mean value to get another definition of efficiency, namely
\begin{eqnarray}
    \label{eff2}
    \eta_S = \frac{ \langle F_R \rangle \langle V \rangle}{\mathcal P_{in}} = \frac{\langle V \rangle^2}{R\mathcal P_{in}} = \frac{\langle V \rangle^2}{\langle V^2 \rangle - \langle V^2 \rangle_{eq}} = \frac{\langle V \rangle^2}{\langle V \rangle^2 + \sigma_V^2 - k_BT/C}. 
\end{eqnarray}
This quantity is called the \emph{Stokes efficiency} \cite{wang2002,schilwa,wang2009}. Let us note that it depends explicitly on mass $C$ of the Brownian particle and and only implicitly on the friction coefficient $R$ via the Langevin equation (\ref{Lang1}). 

It should be mentioned that (\ref{eff2}) is not the rate of the work done by the motor on its surroundings (viscous medium). Moreover, it is not a mean power $\mathcal P_R$ to overcome the friction force which correct form reads 
\begin{equation}
    \label{PR}
 \mathcal P_R = \langle F_R V \rangle = \frac{\langle V^2 \rangle}{R}. 
\end{equation}
However, this expression cannot be put as a numerator in the definition of the efficiency because there are regimes where the mean velocity is extremely small (numerically zero), $\langle V \rangle \approx 0$ but $\langle V^2 \rangle \neq 0$ and the efficiency could be large even though  the particle does not move on average in one direction. It is the main reason why the Stokes efficiency is more adequate is such cases as considered in the paper. 

Another possible definition of the efficiency is based on the remark that what we observe in the long time regime is the average velocity. Therefore we can introduce "kinetic power" as the "kinetic energy" of the particle per the period $T$, 
\begin{equation}
    \label{ek}
    \mathcal P_k = \frac{ C \langle V \rangle^2}{2T}. 
\end{equation}
One should note that it is not exactly proper definition of the kinetic power as it should be proportional to $\langle V^2 \rangle$ instead of $\langle V \rangle^2$. However, we replaced it with the latter for the reason explained before. Nevertheless, it is still a measure of performance of the motor. If the average velocity increases then $\mathcal P_k$ also grows. We can insert it as a numerator in (\ref{eff1}) and then we get the \emph{kinetic efficiency}
\begin{eqnarray}
    \label{eff3}
    \eta_k = \frac{\mathcal P_k}{\mathcal P_{in}} = \frac{R C}{2T} \frac{\langle V \rangle^2}{\langle V^2 \rangle - \langle V^2 \rangle_{eq}} = \frac{\tau_r}{2T}\,\eta_S. 
\end{eqnarray}
This quantifier can be used only when the time-periodic force is switched on. Then qualitatively, it is similar to the Stokes efficiency. However, the dependence on the mass $C$, the friction coefficient $R$ and the period $T$ is different. Both the Stokes efficiency and the kinetic efficiency are consistent with our intuition: a decrease of fluctuations $\sigma_V^2$ leads to a smaller input power and hence to an increase of the efficiency. Consequently, the transport is optimized in regimes that \textit{maximize} the directed velocity and \textit{minimize} its fluctuations. Because the kinetic efficiency is proportional to the Stokes efficiency, below we analyze only the last one.
\section{The results}
\label{results}
\subsection{Dimensionless model}
\label{dimless}
There are several dimensionless forms of Eq. (\ref{Lang1}) in dependence of the choice of scaling time. In this system there are four characteristic frequencies: plasma frequency $\omega_p^2 = 2eJ_1/\hbar C$, the characteristic frequency of the junction $\omega_c = 2eR J_1/\hbar$, the frequency $\omega_r =1/RC$ related to the relaxation time and the frequency $\Omega$ of the ac current. There are three independent characteristic time scales related to these frequencies (note that $\omega_p^2=\omega_c \omega_r$). Here, we follow \cite{zapata1996prl} and define the new phase $x$ and the dimensionless time $\hat{t}$ as
\begin{equation}
	\label{eq28}
	x = \frac{\varphi + \pi}{2}, \quad s = \frac{t}{\tau_c}, \quad \tau_c = \frac{\hbar}{eRJ_l}. 
\end{equation}
Then (\ref{Lang1}) takes the dimensionless form
\begin{equation}
    \label{eq29}
    \tilde C \ddot{x}(s) + \dot{x}(s) = -U'(x(s)) + F + a\cos(\omega s) + \sqrt{2D}\,\hat{\xi}(s),
\end{equation}
where the dot and prime denotes a differentiation over the dimensionless time $s$ and the phase $x$, respectively. We introduced a spatially periodic potential $U(x)$ of period $2\pi$ of the following form \cite{zapata1996prl}
\begin{figure}[t]
    \centering
    \includegraphics[width=0.49\linewidth]{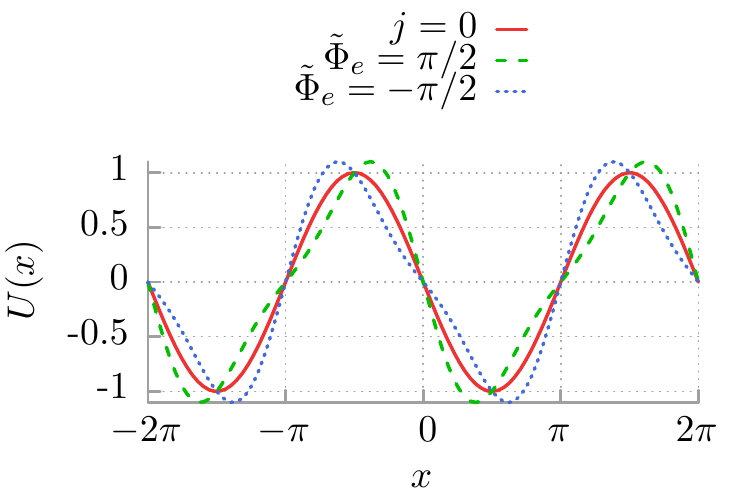}
    \caption{The potential (\ref{eq30}) for the symmetric case $j=0$ (solid red line) in comparison with the ratchet potential for $j=1/2$ and two values of the external magnetic flux $\tilde\Phi_e = \pi/2$ (dashed green line) and $\tilde\Phi_e = -\pi/2$ (dotted blue line).}
    \label{fig2}
\end{figure}
\begin{equation}
    \label{eq30}
    U(x) = - \sin(x) - \frac{j}{2} \sin(2x + \tilde\Phi_e - \pi/2).
\end{equation}
This potential is reflection-symmetric if there exists $x_0$ such that $U(x_0 + x)= U(x_0 - x)$ for any $x$. If $j \neq 0$, it is generally asymmetric and its reflection symmetry is broken, see Fig. \ref{fig2}. We classify this characteristics as a ratchet potential. However, even for $j \neq 0$ there are certain values of the external flux $\tilde \Phi_e$ for which it is still symmetric. The dimensionless capacitance $\tilde C$ is the ratio between two characteristic time scales $\tilde C = \tau_r/\tau_c$, where the relaxation time is $\tau_r = RC$. Other re-scaled parameters are $j = J_2/J_1$, $F = I_0/J_1$, $a = A/J_1$ and $\omega = \Omega\tau_c$. The rescaled zero-mean Gaussian white noise $\hat{\xi}(s)$ has the auto-correlation function \mbox{$\langle \hat{\xi}(s)\hat{\xi}(u) \rangle = \delta(s-u)$} and its intensity $D = e k_B T/\hbar J_1$ is the quotient of the thermal and the Josephson coupling energy. The dimensionless voltage $v(t) = \dot x(s) = V(t)/ RJ_1$ and therefore the physical average voltage $\langle V \rangle$ is given by the relation
\begin{equation}
    \label{eq31}
    \langle V \rangle = R J_1 \langle v \rangle.
\end{equation}
In particular, after such a scaling procedure the dimensionless input power $P_{in}$ is expressed as
\begin{equation} \label{pin}
	 P_{in} = \langle v^2 \rangle - D/\tilde C
\end{equation}
and consequently, the Stokes efficiency reads
\begin{equation}
	\eta_S = \frac{\langle v \rangle^2}{\langle v \rangle^2 + \sigma_v^2 - D/\tilde C} = \frac{\langle v \rangle^2}{\langle v^2 \rangle - D/\tilde C}.
\end{equation} 
The key feature for the occurrence of the directed transport $\langle v \rangle \neq 0$ is the symmetry breaking. This is the case when either the dc current $F \neq 0$ or the reflection symmetry of the potential $U(x)$ is broken.

The system described by (\ref{eq29}) becomes deterministic when the thermal noise intensity $D$ is set to zero. Even in this case it exhibits complex dynamics including chaotic regimes \cite{jung1996, mateos2000}. The application of noise generally smooths out its characteristic response function. There are two classes of states of the driven system dynamics: the locked states, in which the phase stays inside finite number of potential wells and the running states for which it runs over the potential barriers. The latter are crucial for the occurrence of the transport. They can be either chaotic (diffusive) or regular.
\subsection{Details of simulations}
The Fokker-Planck equation (\ref{FP}) corresponding to the Langevin equation (\ref{Lang1}) cannot be solved by use of closed analytical forms. Therefore, in order to obtain the relevant transport characteristics we have to resort to comprehensive numerical simulations of the driven Langevin dynamics. We have integrated (\ref{eq29}) by employing a weak version of the stochastic second order predictor corrector algorithm \cite{platen} with a time step typically set to about $10^{-3} \cdot 2\pi/\omega$. Since (\ref{eq29}) is a second-order differential equation, we have to specify two initial conditions $x(0)$ and $\dot{x}(0)$. Moreover, because for some regimes the system may be non ergodic in order to avoid the dependence of the presented results on the specific selection of initial conditions we have chosen phases $x(0)$ and dimensionless voltages $\dot{x}(0)$ equally distributed over interval $[0, 2\pi]$ and $[-2,2]$, respectively. All quantities of interest were ensemble-averaged over $10^3 - 10^4$ different trajectories which evolved over $10^3 - 10^4$ periods of the external ac driving. Numerical calculations were done by use of a CUDA environment implemented on a modern desktop GPU. This scheme allowed for a speed-up of a factor of the order $10^3$ times as compared to a common present-day CPU method \cite{januszewski2009, spiechowiczcpc}. Part of our so obtained results are presented next.
\begin{figure*}[t]
    \centering
    \includegraphics[width=0.3\linewidth]{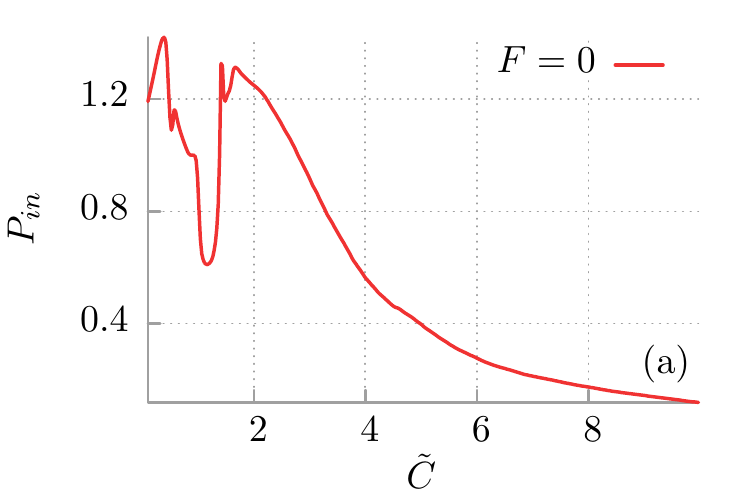}
    \includegraphics[width=0.3\linewidth]{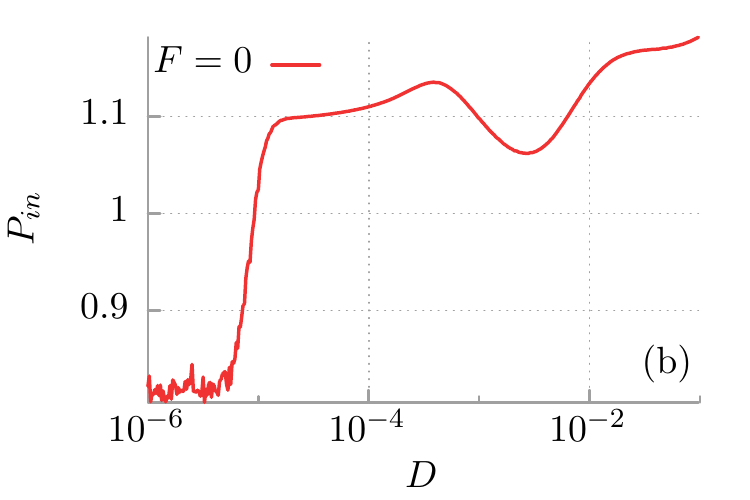}
    \includegraphics[width=0.3\linewidth]{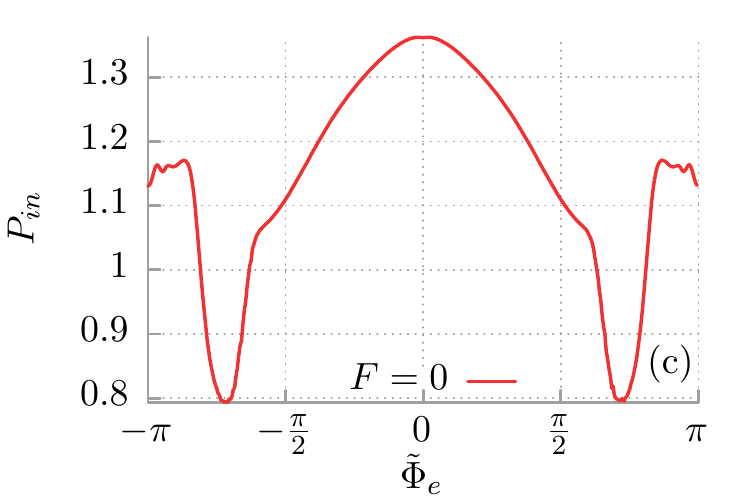}
    \caption{The power (\ref{e2}) delivered by the external current $I(t)$ is presented in the dimensionless form (\ref{pin}) as a function of the dimensionless capacitance $\tilde{C}$, the thermal noise intensity $D$ and the external magnetic flux $\tilde{\Phi}_e$ in panels (a)-(c), respectively. Other parameters read: $a = 1.9$, $\omega = 0.6$, $j = 1/2$. In panel (a): $D = 9.7 \cdot 10^{-5}$ and $\tilde{\Phi}_e = \pi/2$. In panel (b): $\tilde{C} = 0.645$ and $\tilde{\Phi}_e = \pi/2$. In panel (c): $\tilde{C} = 0.645$ and $D = 9.7 \cdot 10^{-5}$.}
    \label{fig3A}
\end{figure*}
\subsection{Power delivered by external current}
Let us begin analysis of the SQUID efficiency by looking at the power $P_{in}$ delivered by the external current $I(t)$. Notably, it depends implicitly not only on the parameters of the applied external current ($F$, $a$, $\omega$) but also on the quantities characterizing the device like the capacitance $\tilde{C}$. We have found that generally the input power (\ref{pin}) tends to increase for larger values of the dc current $F$ and ac driving amplitude $a$. The dependence on the frequency $\omega$ is more complex. However, in most cases $P_{in}$ is relatively large when $\omega$ is small. It is because very fast oscillation of the driving current cannot induce neither the average voltage $\langle v \rangle$ nor $\langle v^2 \rangle$. In panel (a) of Fig. \ref{fig3A} we show  the representative dependence of the input power $P_{in}$ on the dimensionless capacitance $\tilde{C}$ of the SQUID. One can observe that $P_{in}$ is maximal for the overdamped or close to damped regime and decreases when $\tilde{C}$ grows. Since in the mechanical framework the capacitance $\tilde{C}$ translates to the mass of the Brownian particle it is intuitively clear that when the inertial term becomes large then the device needs more power to response equivalently. Perhaps the most surprising is the fact that $P_{in}$ depends explicitly on the thermal noise intensity $D$, i.e. on temperature of the system. Typically, it decreases for increasing $D$. However, there are also regimes for which $P_{in}$ is enhanced by thermal noise. In panel (b) of Fig. \ref{fig3A} we exemplify this situation. Indeed, for a wide interval of temperature the input power is almost monotonically increasing function of the noise intensity $D$. Finally, the influence of the constant external magnetic flux $\tilde{\Phi}_e$ on $P_{in}$ is depicted in the last panel. It is remarkable that one can tune the input power $P_{in}$ by changing the external magnetic flux. The reader should note that for the presented regime it is maximal when $\tilde{\Phi}_e = 0$, i.e. potential $U(x)$ is reflection symmetric. In such a case there is no average voltage drop $\langle v \rangle = 0$ across the device when additionally the dc current $F$ vanishes. It follows that large input power $P_{in}$ does not necessarily translate into the efficient directed transport.
\begin{figure}[t]
    \centering
    \includegraphics[width=0.49\linewidth]{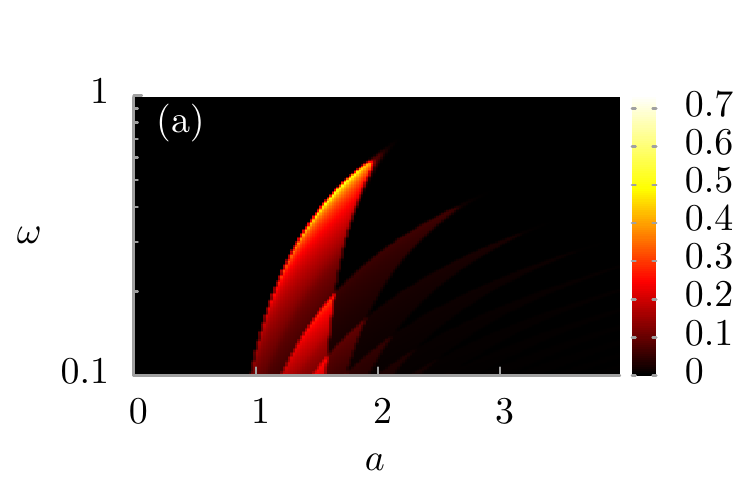} 
    \includegraphics[width=0.49\linewidth]{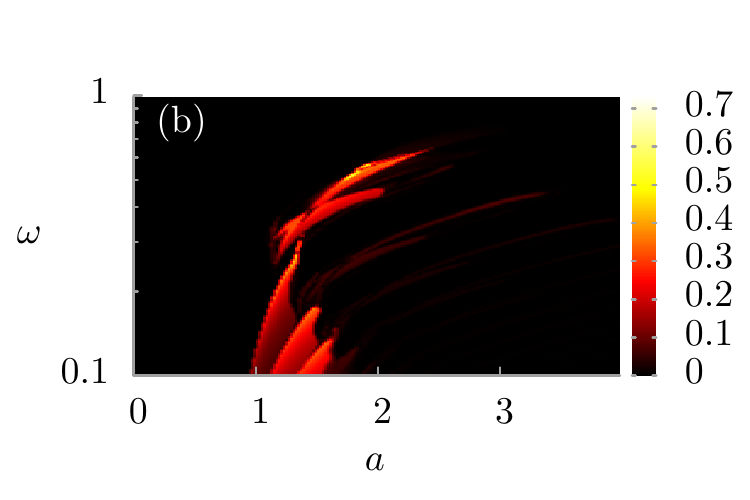} \\
    \includegraphics[width=0.49\linewidth]{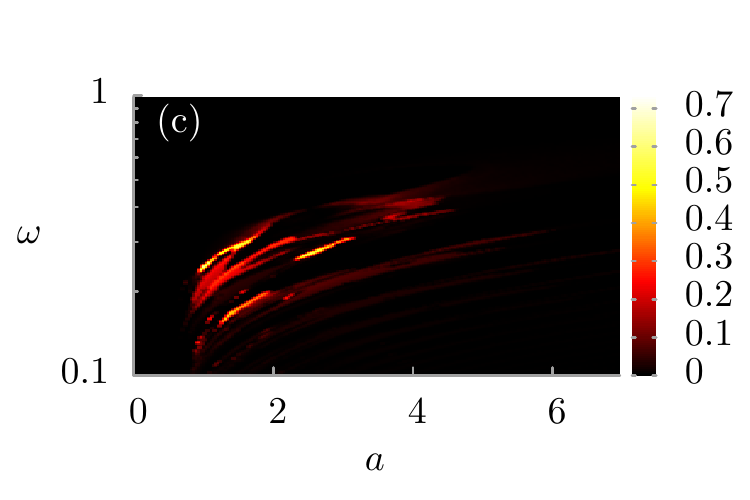}
    \caption{The Stokes efficiency $\eta_S$ defined by Eq. (\ref{eff2}) in the parameter plane $\{a, \omega\}$ of the ac current for three distinct regimes: overdamped ($\tilde{C} = 0.2$), moderate damping ($\tilde{C} = 2$) and underdamped ($\tilde{C} = 10$) in panels (a), (b) and (c), respectively. The remaining parameters are: $F = 0$, $D = 10^{-5}$, $j = 1/2$ and $\tilde{\Phi}_e = \pi/2$.}
	\label{fig3}
\end{figure}
\subsection{Tailoring Stokes efficiency}
\label{Stokes}
The system described by Eq. (\ref{eq29}) has a 7-dimensional parameter space $\{ \tilde{C}, a, \omega, F, j, \tilde{\Phi}_e, D\}$. We set $F = 0$ and check how it depends on the remaining system parameters. We limit our considerations to positive $a$ because the system (\ref{eq29}) is symmetric under changes of sign of $a$. Depending on the magnitude of the dimensionless capacitance $\tilde{C}$ of the device it can operate in three distinct regimes: overdamped ($\tilde{C} \to 0$), damped (moderate $\tilde{C}$) and underdamped ($\tilde{C} \to \infty$). We note that the conditions that are necessary for the generation and control of the direction of transport have been extensively studied in these regimes in our recent work \cite{spiechowicz2014}. Since very fast oscillation of the driving current cannot induce the average voltage $\langle v \rangle$ it is sufficient to limit our considerations to low and moderate ac driving frequencies $\omega$. We have performed scans of the following area of the parameter space $\tilde{C} \times a \times \omega \in [0.1;10] \times [0;10] \times [0.1;1]$ at a resolution of 200 points per dimension to determine the general behavior of the system. The results are depicted in Fig. \ref{fig3}.

We can see that regardless of the regime in which the device operates its Stokes efficiency $\eta_S$ is zero or negligibly small for $a < 1$ and high frequencies $\omega$. This is due to the fact that the rocking mechanism is either too weak or too fast to induce finite average voltage $\langle v \rangle$. The areas of non-zero efficiency $\eta_S$ have a striped structure. For a given amplitude $a$, the ratchet behavior generally tends to disappear as the frequency $\omega$ grows. On the other hand, for a given frequency, there is optimum amplitude $a$ that maximize the Stokes efficiency. The increase of the capacitance $\tilde{C}$ causes blurring of the regions for which the efficiency is nonzero. Moreover, this tendency is often accompanied by its reduction. Consequently, the studied device operates best in the overdamped or close to damped regimes.
\begin{figure*}[t]
    \centering
    \includegraphics[width=0.3\linewidth]{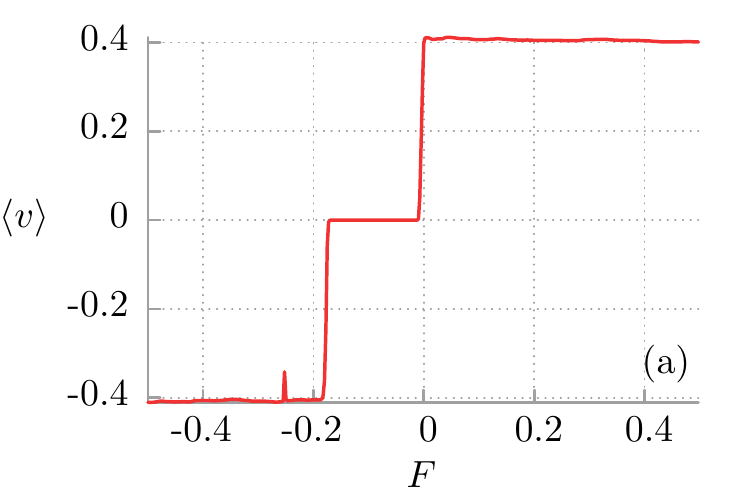}
    \includegraphics[width=0.3\linewidth]{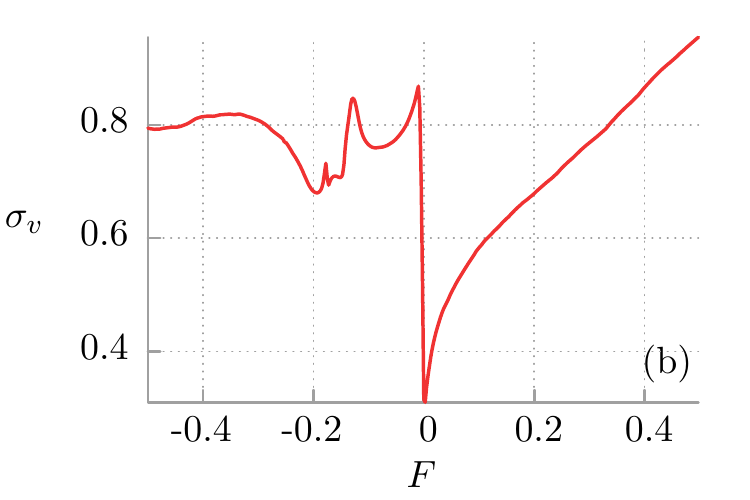}
    \includegraphics[width=0.3\linewidth]{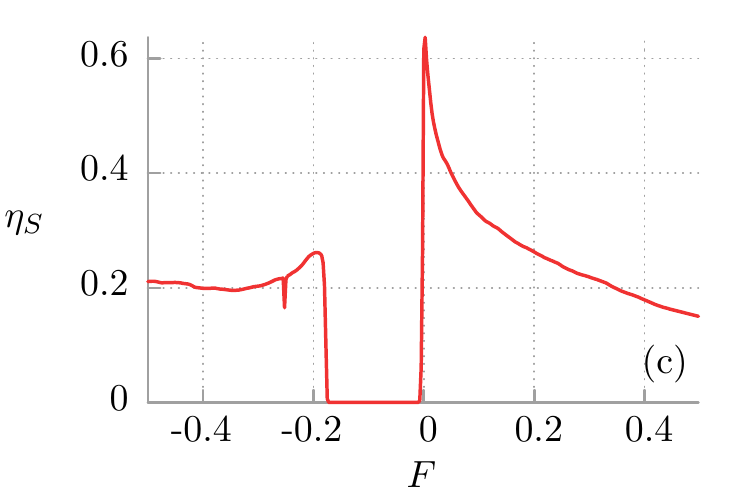} \\
    \includegraphics[width=0.3\linewidth]{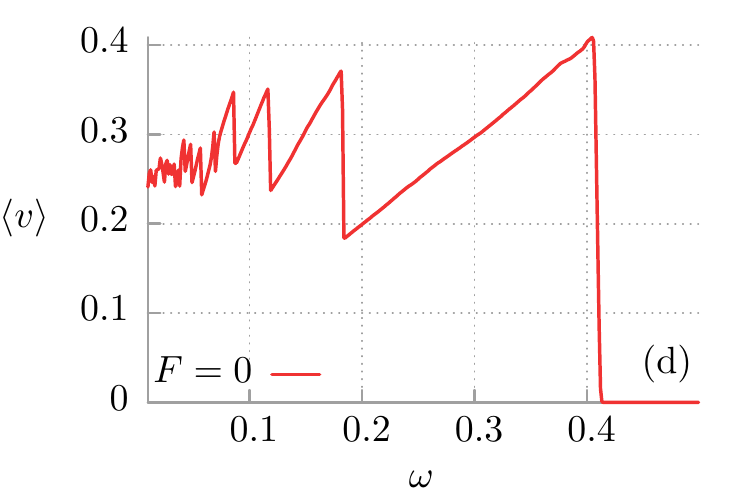}
    \includegraphics[width=0.3\linewidth]{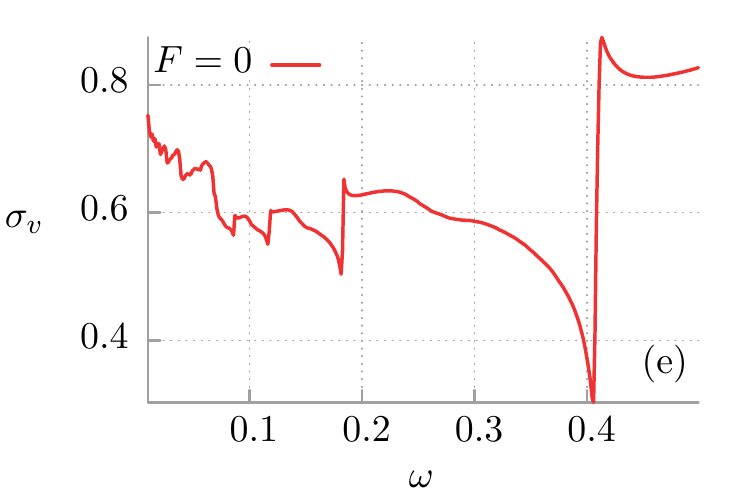}
    \includegraphics[width=0.3\linewidth]{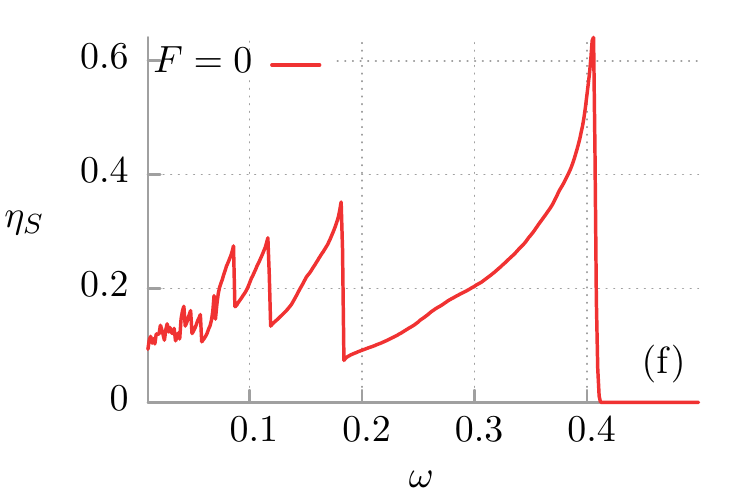} \\
    \includegraphics[width=0.3\linewidth]{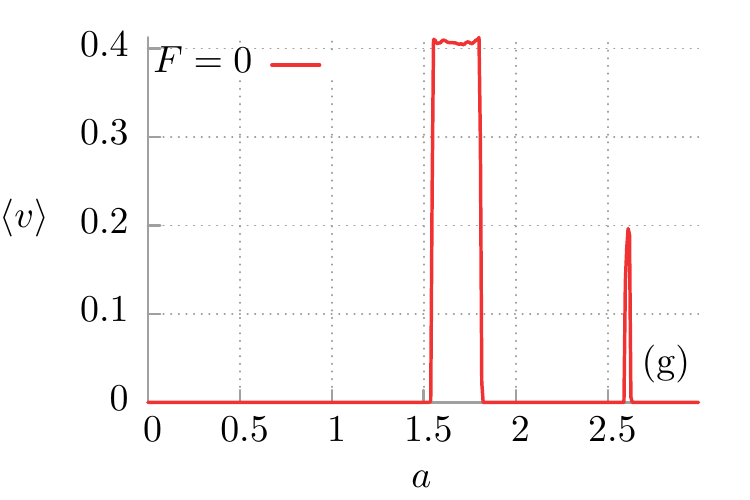}
    \includegraphics[width=0.3\linewidth]{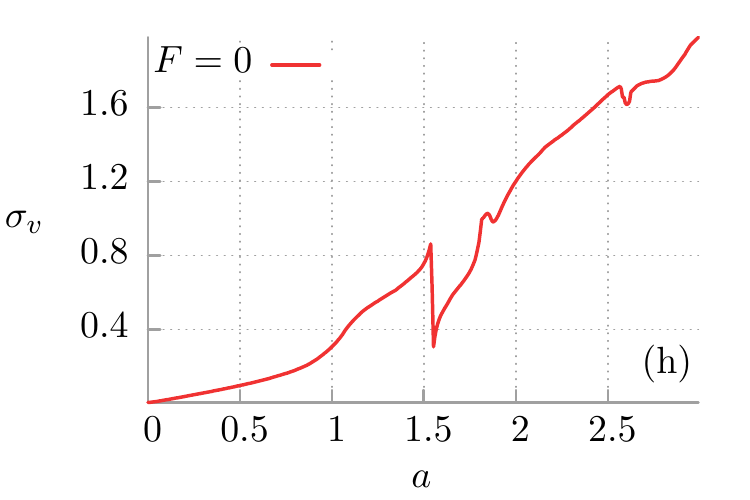}
    \includegraphics[width=0.3\linewidth]{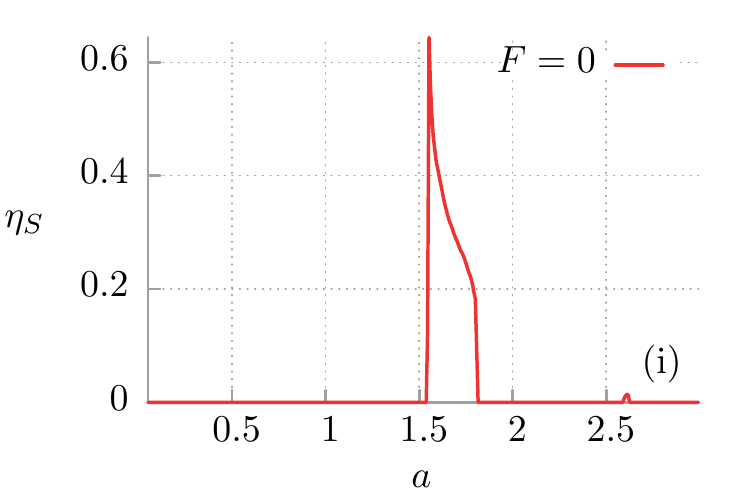} \\
    \includegraphics[width=0.3\linewidth]{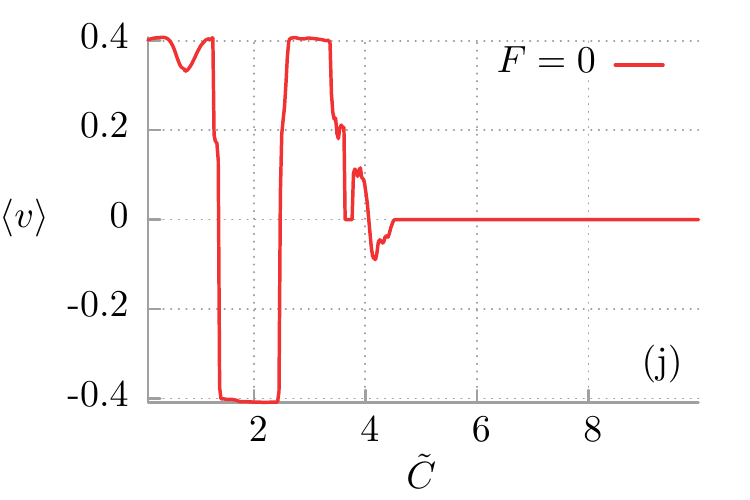}
    \includegraphics[width=0.3\linewidth]{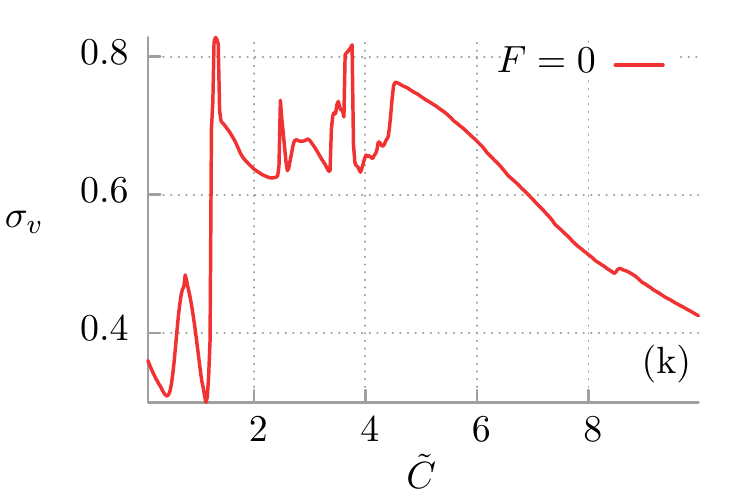}
    \includegraphics[width=0.3\linewidth]{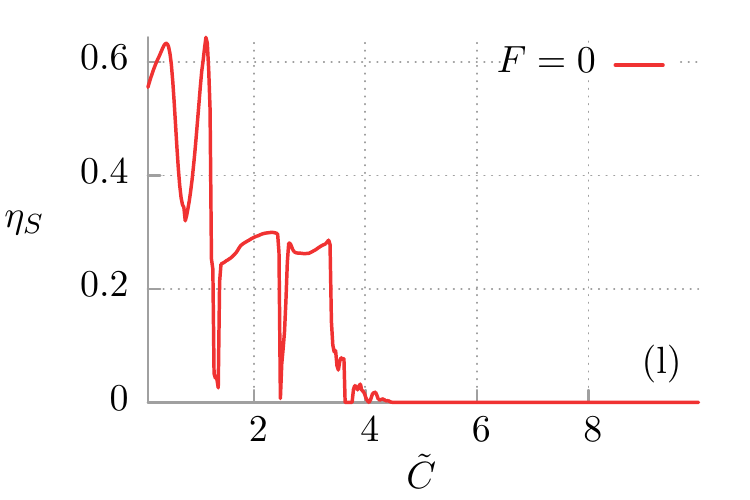} \\
    \includegraphics[width=0.3\linewidth]{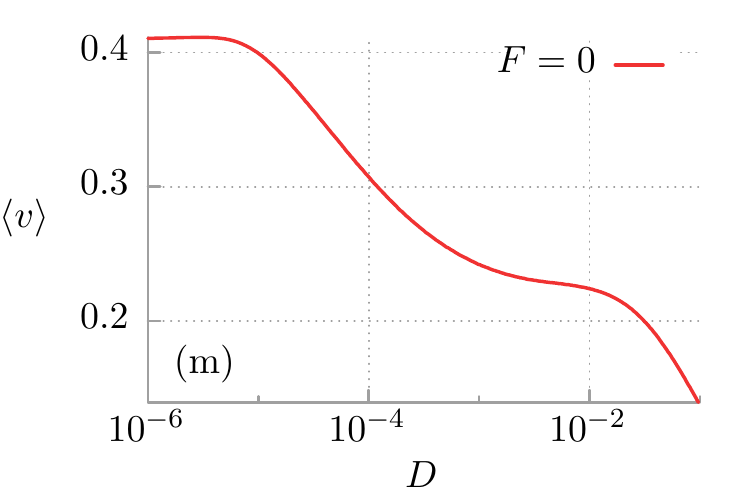}
    \includegraphics[width=0.3\linewidth]{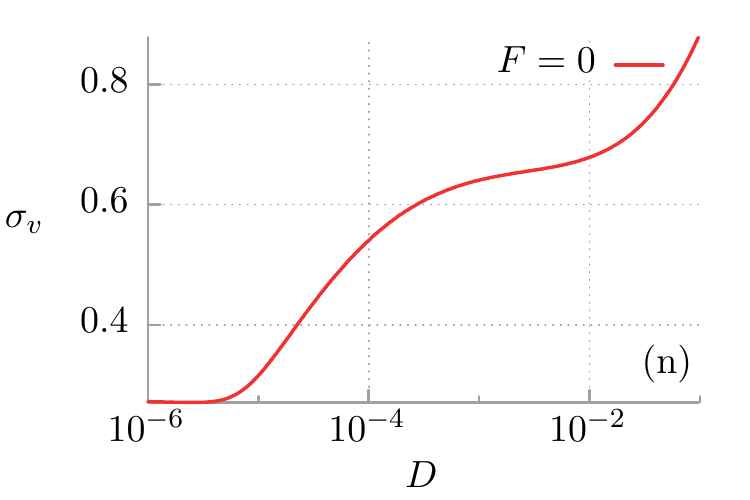}
    \includegraphics[width=0.3\linewidth]{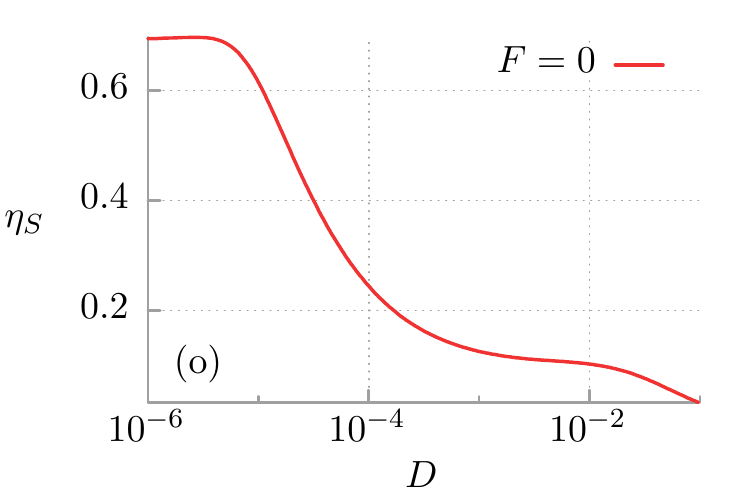} \\
    \caption{Optimal regime for the Stokes efficiency of the transport occurring in (\ref{eq29}). The dependence of the average voltage $\langle v \rangle$, its fluctuations $\sigma_v$ and finally the Stokes efficiency $\eta_S$ on the static dc-bias $F$, frequency $\omega$, amplitude $a$, capacitance $\tilde{C}$ and thermal noise intensity $D$ is presented in panels (a)-(o). Parameters are: $F = 0$, $\omega = 0.406$, $a = 1.55$, $\tilde{C} = 0.496$, $D = 10^{-5}$, $j = 1/2$ and $\tilde{\Phi}_e = \pi/2$.}
    \label{fig4}
\end{figure*}
\subsection{Optimal regime}
\label{optimal}
We have explored the parameter space of the system (\ref{eq29}) and we have been able to detect a regime for which the efficiency $\eta_S$ is {\it globally maximal}. It is in the vicinity of the point 
$\{\tilde C, a, \omega\}= \{0.496, 1.55, 0.406\}$. 
All relevant transport characteristics, i.e. the average voltage $\langle v \rangle$, its fluctuations $\sigma_v$ and the efficiency $\eta_S$ corresponding to neighborhood of this set of parameters are presented in Fig. 5. 

Let us begin with the dependence of the transport characteristics on the dc current $F$. It is depicted in Fig. \ref{fig4}(a)-(c) for small values of $F \in(-0.5, 0.5)$. Panel (a) presents the current-voltage curve. In the low temperature limit ($D = 10^{-5}$) the average voltage is almost quantized at values $n\omega, \, n = 0, \pm 1, ..$. For a symmetric potential, these plateaus correspond to standard Shapiro steps \cite{baronepaterno}. However, in our case also steps at half integer multiplies of $\omega$ can be observed. This is due to the deviation of $U(x)$ from a simple $\sin{x}$ form, which is the sole case for which steps lie only at integer values of $n\omega$ \cite{zapata1996prl}. However, in both the symmetric and asymmetric cases a proper amount of noise is sufficient to wipe out their evident structure \cite{kautz1996}. This Shapiro-like current-voltage curve is characteristic for the device operating in the low temperature limit of overdamped or damped regimes. Panel (b) of the same Fig. 5 presents the dependence of the voltage fluctuations $\sigma_v$ on the dc current $F$. It is rather complicated non-linear and non-monotonic function of $F$ without any immediately obvious relation to the average voltage of panel (a). However, the most important observation is that the voltage fluctuations are minimal for $F=0$. This fact is of fundamental importance for the influence of $F$ on $\eta_S$. In Fig. \ref{fig4}(c) we can see that $\eta_S$ is locally maximal for $F=0$. For large values of $F$ (not shown here) the mean voltage is an almost linear function of $F$ and the efficiency approaches the value 1. It support the statement in Ref. \cite{schilwa} that when the Stokes efficiency is close to $1$, the driving resembles the constant force. 

The role of the frequency $\omega$ of the ac current is illustrated in Fig. \ref{fig4}(d)-(f). Panel (d) presents the dependence of the average voltage $\langle v \rangle$ on $\omega$. In the adiabatic limit $\omega \to 0$ it undergoes rapid oscillations. This behavior has its reflection in the influence of the frequency on the voltage fluctuations $\sigma_v$ (see Fig. \ref{fig4}(e)). According to the previous statement very fast oscillations of the driving current cannot induce the non-zero average voltage. Therefore for sufficiently high frequency there is no transport and as a consequence the Stokes efficiency $\eta_S$ is zero. However, a strong peak of efficiency is observed for moderate value of $\omega = 0.406$. It is associated with the fact that for this frequency the average voltage $\langle v \rangle$ is maximal and simultaneously its fluctuations $\sigma_v$ are minimal.

The impact of the amplitude $a$ of the ac current is shown in Fig. \ref{fig4}(g)-(i), respectively. In particular, the resonance-like behavior is observed in the dependence of the average voltage $\langle v \rangle$ on the amplitude $a$ (see Fig. \ref{fig4}(g)). Apart from two clearly visible peaks there is almost imperceptible small directed transport. This fact has critical impact on the functional dependence of the efficiency $\eta_S$. It is proportional to $\langle v \rangle^2$ so it vanishes too when the device response is zero. The influence of variation of the amplitude $a$ on the voltage fluctuations $\sigma_v$ is depicted in panel (h). It is almost linearly increasing function of $a$. Only one evident deviation from this trend can be noted, i.e. a local minimum around $a = 1.55$ corresponding to the first high peak in Fig. \ref{fig4}(g). It should be stressed that there is no any contradiction with the dependence of the average voltage since $\langle v \rangle = 0$ does not necessarily mean $\langle v^2 \rangle = 0$ and therefore $\sigma_v$ can at the same time assume nonzero value. 

The dependence of all relevant transport characteristics on the capacitance $\tilde{C}$ is very complicated as shown in Fig. \ref{fig4}(j)-(l). The first panel of this group shows the average voltage versus the capacitance $\tilde{C}$. We note the important feature of the voltage reversal \cite{jung1996, machura2007}: starting from zero, the voltage changes its sign from positive to negative and again in the opposite direction as $\tilde{C}$ grows. Therefore the capacitance can serve as a parameter to manipulate the direction of transport processes. The efficiency $\eta_S$ is maximal close to the border between the overdamped and damped regimes. It is (almost) zero in the underdamped limit which corresponds to large capacitance $\tilde{C} \to \infty$. This is a consequence of the fact that for this regime the average voltage $\langle v \rangle$ vanishes or is negligibly small.

The last three panels of Fig. \ref{fig4} depict the influence of the thermal noise intensity $D$ on all previously studied quantities. An increase of the noise intensity $D$ leads to both monotonic decrease of the induced average voltage $\langle v \rangle$ and increase of its fluctuations $\sigma_v$. Consequently, the efficiency is the best in the low temperature regime when the deterministic dynamics of the system (\ref{eq29}) plays a crucial role.
\begin{figure*}[t]
    \centering
    \includegraphics[width=0.3\linewidth]{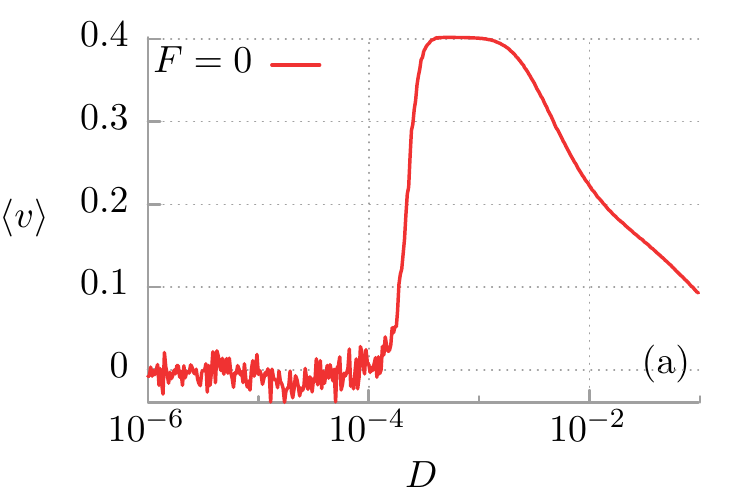}
    \includegraphics[width=0.3\linewidth]{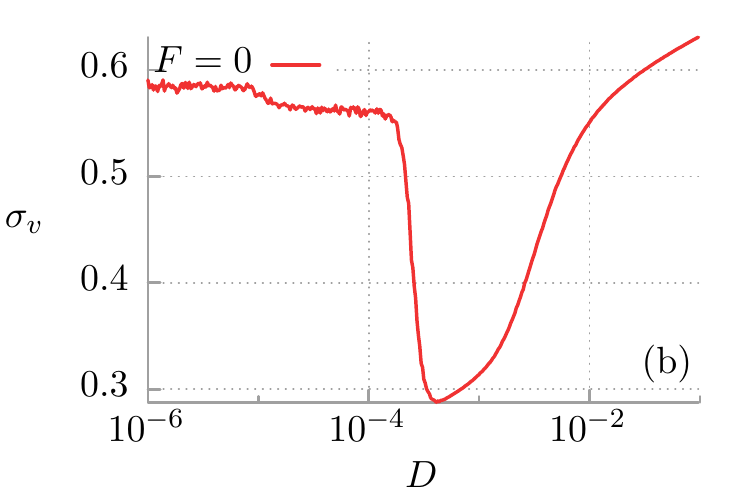}
    \includegraphics[width=0.3\linewidth]{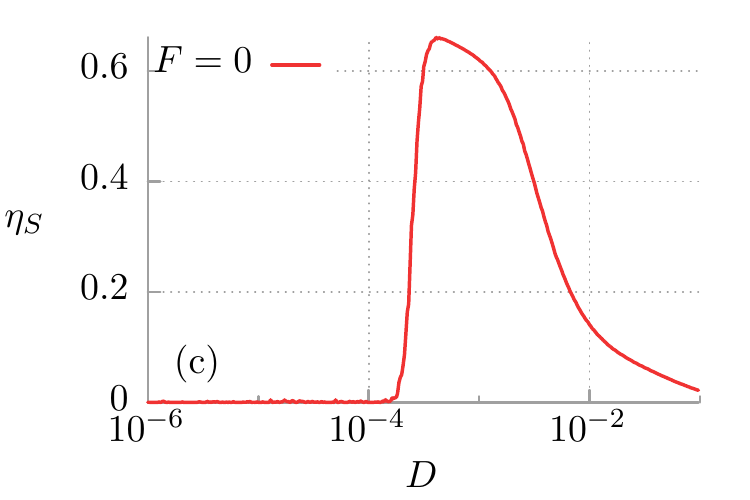}
    \caption{Noise enhanced Stokes efficiency. The dependence of the average voltage $\langle v \rangle$, its fluctuations $\sigma_v$ and finally the efficiency $\eta_S$ on thermal noise intensity $D$ is presented in panels (a)-(c), respectively. Parameters read: $a = 1.899$, $\omega = 0.403$, $\tilde{C} = 6$, $j = 1/2$ and $\tilde{\Phi}_e = \pi/2$.}
    \label{fig5}
\end{figure*}
\subsection{Noise enhanced Stokes efficiency}
\label{noise}
We have found the opposite scenario when thermal noise enhances the efficiency. This perhaps surprising effect is exemplified in Fig. \ref{fig5}. Panel (c) presents the dependence of the efficiency $\eta_S$ on $D$. Evidently, in some intervals of $D$, the increase of temperature causes the increase of $\eta_S$. There is also optimal value of temperature or equivalently the thermal noise intensity $D \approx 0.0004$ for which the efficiency takes its maximum. Moreover, in this case the ratchet mechanism is solely activated by thermal equilibrium fluctuations as for low noise intensity no rectification can be observed. This statement is confirmed in the functional dependence of the average voltage $\langle v \rangle$ which is presented in panel (a). It is also remarkable that in this regime an increase of thermal noise intensity $D$ leads to a decrease of voltage fluctuations $\sigma_v$.
\begin{figure*}[t]
    \centering
    \includegraphics[width=0.3\linewidth]{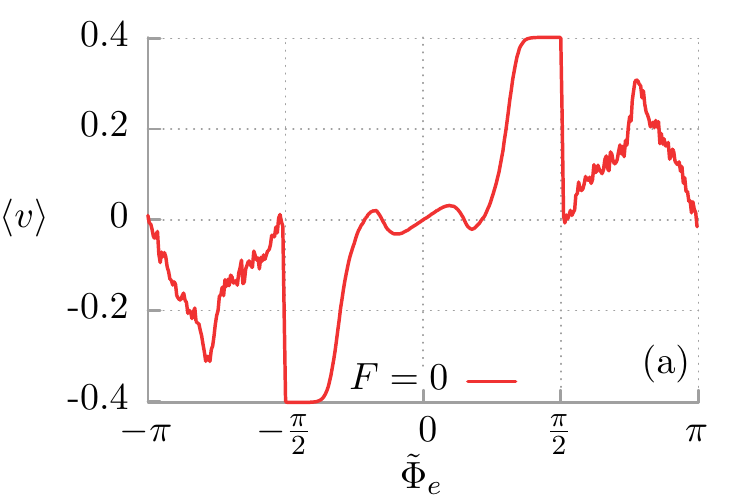}
    \includegraphics[width=0.3\linewidth]{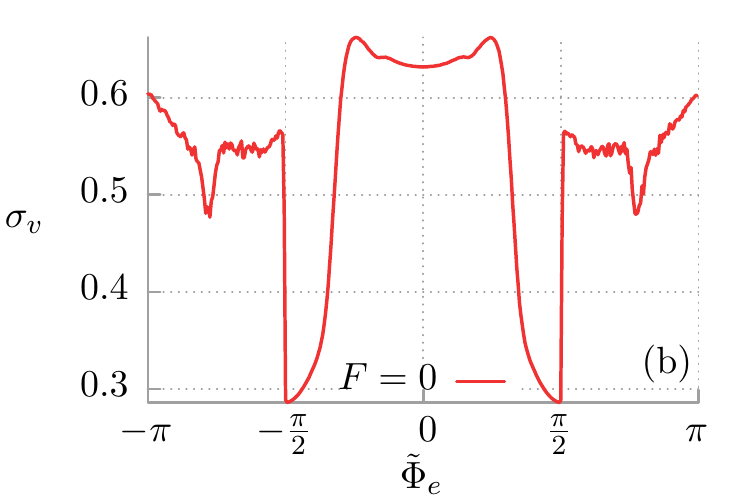}
    \includegraphics[width=0.3\linewidth]{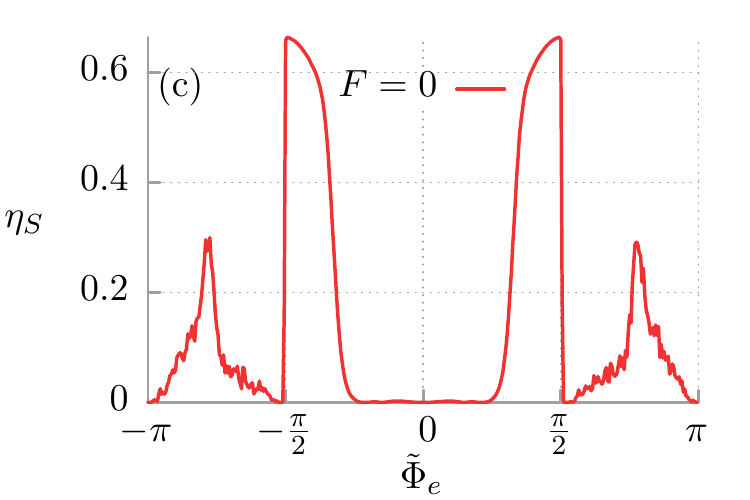}
    \caption{Impact of the external magnetic flux on the average voltage $\langle v \rangle$, its fluctuations $\sigma_v$ and finally the  efficiency $\eta_S$ is presented in panels (a)-(c) for the noise intensity $D=0.0004$ which corresponds to the maximum in Fig. 6(c). Other parameters are the same as in Fig. \ref{fig5}.}
    \label{fig6}
\end{figure*} 
\subsection{Impact of external magnetic flux}
\label{flux}
As it was shown before, the efficiency can be tuned in several ways. However, it seems that from the experimental point of view the simplest method is to vary the external constant magnetic flux $\tilde{\Phi}_e$. The dependence of the average voltage $\langle v \rangle$, its fluctuations $\sigma_v$ and the efficiency on the external magnetic flux $\tilde{\Phi}_e$ in the previously presented regime for which thermal noise induces the ratchet effect (cf. Fig. \ref{fig5}) is depicted in Fig. \ref{fig6}. From the symmetry considerations of (\ref{eq29}) it follows that for an arbitrary integer number $n$, the transformation $\tilde{\Phi}_e \rightarrow 2\pi n - \tilde{\Phi}_e$ reverses the sign of the average voltage $\langle v \rangle \rightarrow -\langle v \rangle$. This fact can be directly observed in panel (a) of Fig. \ref{fig6}. However, this is not the case for the voltage fluctuations $\sigma_v$, where they are symmetric around $\tilde{\Phi}_e = 0$. A careful inspection of panel (b) reveals that one can reduce magnitude of $\sigma_v$ by nearly two times just by correct adjustment of the external magnetic flux. This fact has further consequences in the dependence of the efficiency which is depicted in panel (c). It can be slightly tuned by a small variation of the external magnetic flux. 

\begin{figure}[t]
    \centering
	\includegraphics[width=0.49\linewidth]{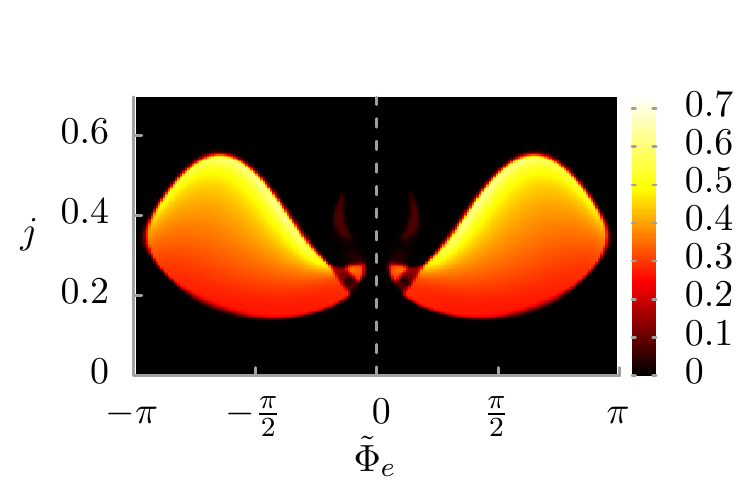}
	\includegraphics[width=0.49\linewidth]{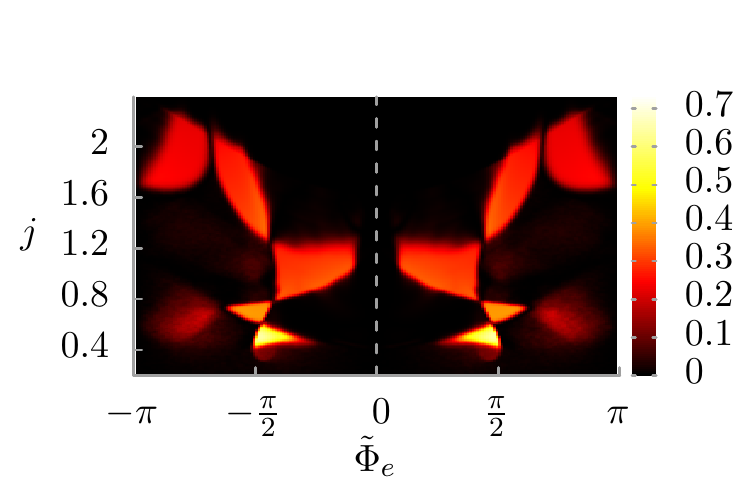}
    \caption{The Stokes efficiency of the rocked SQUID in the parameter plane $\{\tilde{\Phi}_e, j\}$. The left panel corresponds to the optimal regime as in Fig. \ref{fig4} and the remaining parameters are the same as there. The right panel presents the regime depicted in Fig. \ref{fig5} with $D=0.0004$.}
    \label{fig7}
\end{figure}

In Fig. \ref{fig7} we present how the Stokes efficiency behaves in the parameter plane $\{\tilde{\Phi}_e, j\}$ that specifies the form of the spatially periodic potential $U(x)$. For both sufficiently small and large $j$ it vanishes completely. One can observe that for a given external magnetic flux $\tilde{\Phi}_e$ the Stokes efficiency generally tends to increase as the parameter $j$ grows. On the contrary, for a given $j$ there is an optimal value of the external magnetic flux $\tilde{\Phi}_e$ for which the Stokes efficiency is maximal. We note that for two presented regimes, the set structure of the non-zero efficiency in the parameters plane $\{\tilde{\Phi}_e, j\}$ is radically different. The left panel looks like butterfly wings and the right is similar to a rocking horse.
\section{Summary}
\label{summary}
In this paper, we comprehensively studied the Stokes efficiency of the asymmetric SQUID  in the case of the non-zero capacitance of all Josephson junctions and in presence of thermal noise. It allowed to analyze transport properties in the system for the entire scale of regimes: starting from overdamped, by damped and finally underdamped one. We focused on the connection between the directed transport characterized by the voltage across the SQUID  and its efficiency.  In particular, we examined voltage fluctuations and energetic performance of the device. 
We derived the expression for the power delivered by externally applied current and discuss its dependence on the system parameters. Apart from the expected influence of the current parameters $I_0$, $A$ and $\Omega$ it also depends on the thermal noise intensity $D$, i.e. on temperature of the system. 

We have found that regions of low efficiency of the SQUID dominates in the parameter space. However, we have identified remarkable and distinct regimes of high efficiency $\eta_S \approx 0.65$. It turns out that the device operates best in the overdamped or close to damped regimes. Moreover, with the help of the computational power of modern GPU supercomputers we have identified the tailored set of parameters for which the efficiency $\eta_S$ is globally maximal and for this regime we discussed impact of variation of almost all system parameters on the relevant transport quantities. In particular, it follows that thermal fluctuations often have destructive influence on the energetic performance of the device. Moreover, we were able to detect also the regime for which thermal noise enhances the efficiency by inducing the large average voltage and minimizing its variance. Last but not least, we discussed in detail the impact of the external magnetic flux $\tilde{\Phi}_e$ on the performance and effectiveness of the SQUID.

Our results can readily be experimentally verified with an accessible setup consisting of three resistively and capacitively shunted Josephson junctions formed in an asymmetric SQUID device. Some partial transport characteristics like voltage have been experimentally studied in the overdamped regime \cite{sterck2005,sterck2009}. However, the underdamped regime has not been tested and the efficiency has not been measured which makes our study a challenge for experimentalists.

\ack
This work was supported in part by the MNiSW program ”Diamond Grant” (J. S.) and NCN Grant DEC-2013/09/B/ST3/01659 (J. {\L}.).

\end{document}